\documentstyle[12pt]{article}

\begin{document}
\begin{center}

 {\Large {\bf Closed universes can satisfy the holographic principle
 in three dimensions}}
  \vskip 0.8truecm {\large Norman Cruz}
 \vskip 0.2cm {{\it Departamento de F\'{\i}sica, Facultad de
 Ciencia,\\
 Universidad de Santiago de Chile, Casilla 307, Correo 2, Santiago, Chile}  \\
 \small E-mail: {\tt ncruz@lauca.usach.cl }} \vspace{0.2in}

{\large Samuel Lepe} \vskip 0.2cm {{\it Departamento de
F\'{\i}sica, Facultad de  Ciencia,\\
 Universidad de Santiago de Chile, Casilla 307, Correo 2, Santiago, Chile.\\
 Instituto de F\'{\i}sica, Universidad Cat\'{o}lica de Valpara\'{\i}so,
 Casilla 4059, Valpara\'{\i}so, Chile}  \\
 \small E-mail: {\tt slepe@lauca.usach.cl}} \vskip 0.5truecm
\today
\end{center}

\vskip 0.5truecm

\begin{abstract} We examine in details Friedmann-Robertson-Walker
models in $2+1$ dimensions in order to investigate the cosmic
holographic principle suggested by Fischler and Susskind. Our
results are rigorously derived differing from the previous one
found by Wang and Abdalla. We discuss the erroneous assumptions
done in this work. The matter content of the models is composed of
a perfect fluid, with a $\gamma$-law equation of state. We found
that closed universes satisfy the holographic principle only for
exotic matter with a negative pressure. We also analyze the case
of a collapsing flat universe.

\end{abstract}

%%%%%%%%%%%%%%%%%%%%%%%%%%%%%%%%%%%%%%%%%%%%%%%%%%%%%%%%%%%%%%%%%%%
%  Introduccionn
%
%%%%%%%%%%%%%%%%%%%%%%%%%%%%%%%%%%%%%%%%%%%%%%%%%%%%%%%%%%%%%%%%%%%

To achieve a consistent theory of quantum gravity has been one of
the most outstanding challenges in  contemporary physics.  One of
the new principles that emerges as a guideline to this ultimate
theory is the holographic principle (HP). In  simple words HP
establish that all degrees of freedom of a region of space in are
the same as that of a system of binary degrees of freedom
distributed on the boundary of the region \cite{Susskind}. In
addition, the number of  degrees of freedom per unit area must be
no greater than 1 per Planck area. This implies a bound on the
entropy of a region, which must not exceed its area in Planck
units. (For a recent review  of HP , see \cite{Bigatti}).

HP was inspired by the result about the total entropy of matter,
$S_{m}$, inside a black hole. According the black hole physics
$S_{m}= S_{BH} = A/4$, where  $S_{BH}$ is the Bekenstein-Hawking
entropy and $A$ is the area of the event horizon in Planck units
\cite{Bekenstein}. In view of the correspondence between
information and entropy, this result can be interpreted in terms
that all the information about the interior of a black hole is
stored on its horizon. An important number of investigations have
been realized in relation with the use of HP in string theory,
quantum gravity and M-theory. More precisely, the relevance of HP
is supported by the concept of matrix theory which suggest a
holographic nature \cite{Banks} and even more by the discovery
of  the correspondence between string theory in $AdS(5)\bigotimes
S(5)$ and Super Yang Mills theory on the boundary
\cite{Maldacena}.

A specific realization of a general cosmological HP was suggested
by Fischler and Susskind (FS) \cite{Fischler}. Since the
Robertson-Walker models have no boundaries, this formulation
consider the particle horizon as a boundary,  pointing out that
for an adiabatic evolution,  the total entropy of matter within
the particle horizon must be smaller than the area of the horizon,
$S_{m}<A$.

Despite its non triviality, this version of HP has been
successful in its applying to cosmology in the following points :
a) is valid for flat universe filled with matter satisfying all
possible equations of state , under the condition $0 \leq p \leq
\rho $; b) if the holographic bound was satisfied at the Planck
time, later on it will be satisfied even better. Nevertheless, the
FS-formulation is violated in closed universe and in open, closed
and flat universes with a negative cosmological constant
\cite{Kaloper}. These results led to another formulations of HP.
Bousso generalized the entropy bound proposed in the
FS-formulation to include general geometries \cite{Bousso}. Bak
and Rey \cite{Bak} proposed a HP based on the concept of apparent
horizon; Easther and Lowe \cite{Easther} replace the holographic
bound by the requirement that physics obey the generalized second
law of thermodynamics.  Within these proposal, closed universes
satisfy HP.  Other studies found that HP according to  the
FS-formulation is satisfied for a closed universe filled with two
fluids, where one of them has an equation of state, $p= w\rho$,
with $w < -1/3$ \cite{Rama}. "Quintessence" models (QCDM), which
have been considered in order to explain that our universe is
accelerating \cite{Ratra}, contain these type of fluids
\cite{Ratra}.

The first investigations of HP formulated in \cite{Fischler} for
three dimensional gravity were realized by Wang and Abdalla
\cite{Wang}. They studied Friedmann-Robertson-Walker (FRW) models
filled with one type of matter component obeying a $\gamma$-law
equation of state , with $\gamma > 1$. The authors mention that
do not exist classical solution when the universe is filled with a
fluid with negative pressure. They found that HP is violated in
closed universe, despite the sign of the pressure matter.

In this paper, we study FRW universes filled with perfect fluids
obeying a $\gamma$-law equation of state. We include a
generalization of previous investigation for the flat case,
considering that the matter content is represented by a
energy-momentum tensor satisfying the dominant energy condition
(DEC), i. e., $-\rho \leq p \leq \rho$, with $\rho>0$. Our
results require to impose adequate initial conditions consistent
with the quantum nature involved in HP. We found that closed
models filled with only one fluid with negative pressure satisfy
the holographic bound derived from HP formulated in
\cite{Fischler}. We also discuss closed universes with two fluids
which satisfy HP. Due to the importance of AdS spaces in
holography, we discuss the behavior of universes which contain a
negative cosmological constant.

A  homogeneous and isotropic universe in three dimensions is
described by the line element
\begin{equation}
ds^2= -dt^2 + a^2(t) \left ( d\chi^2
     + f_{\kappa}^2(\chi) d\phi^2 \right ),
\label{RW1}
\end{equation}
where $a(t)$ is the scale factor and $f_{\kappa}(\chi) = \sin
\chi, \chi, \sinh \chi$ if the universe is closed ($\kappa=1$),
flat ($\kappa= 0$) or open ($\kappa=-1$)respectively; $0 \leq \chi
< \pi$, for $\kappa=1$, and $0 \leq \chi < \infty$, for
$\kappa=0,-1$. The angle of azimuth satisfy $0 \leq \phi \leq 2
\pi$.

 The Einstein's field equations of this model are
\begin{eqnarray}
(\frac{\dot{a}}{a})^{2} & = & - \frac{k}{a^{2}}+ 2 \pi G \rho ,
 \label{fe1} \\ \frac{\ddot{a}}{a} & = & - 2 \pi G p ,\label{fe2}
\end{eqnarray}
where $G$ is the gravitational constant in $2+1$ dimensions. We
consider that the pressure $p$ and the density $\rho$ of the fluid
are related by the $\gamma$-law
\begin{equation}
p= (\gamma-1) \rho,   \qquad \mbox{with} \qquad 0 \leq \gamma
\leq 2 \,.
 \label{se}
\end{equation}
This means that the energy-momentum tensor satisfies DEC.

The conservation of the energy-momentum tensor  yields
\begin{equation}
\dot{\rho} + 2\frac{\dot{a}}{a}(\rho + p) = 0, \label{conserva}
\end{equation}
where the dot means derivative with respect to $t$. Integrating
(\ref{conserva}) with the equation of state (\ref{se}) we obtain
\begin{equation}
\rho= \rho_{0} \left( \frac{a_0}{a} \right)^{2 \gamma},
\label{density}
\end{equation}
where the constants $a_0$ and $\rho_{0}$ are the scale factor and
the energy density of the universe at the initial time,
respectively. In the following, before to apply the FS's
condition to the $2+1$ dimensional models, we discuss which is the
appropriate initial time.

In $3+1$ dimensions, if $0 \leq \gamma \leq 2/3$, the particle
horizon diverges and the problem is solved defining the particle
horizon as an integral not from $t=0$, but from the Planck time
$t=1$ \cite{Kaloper}. In $2+1$ dimensions, the problem with the
initial condition appears for the simple case of dust, where the
particle horizon is well defined, but $S/A$ is zero at $t=0$.
With these conditions HP cannot be realized for flat universes,
contrary to the interpretation made in \cite{Wang}. In $2+1$
dimensions, quantum gravity effects are important at the Planck
scale, $l_{P}$, which is given by $G\hbar$. The associated Planck
time, $t_{P}$, is directly $G\hbar$ ($c=1$). The presence of
$\hbar$ in $l_{P}$ and $t_{P}$ ensures that the quantum nature of
these scales. Nevertheless, in $2+1$ dimensions, the Planck mass,
$m_{P}$, which can be evaluated from the relation $\triangle E
\triangle t \sim \hbar$, is equal to $1/G$. In order to impose
adequate initial conditions, we assume that $a=1=l_{P}$ at the
Planck time $t=1$. Here and after we will use $G=\hbar=1$. We use
this conditions to obtain first $\rho$ as a function of the scale
factor. Integrating Eq. (\ref{conserva}) with the equation of
state (\ref{se}) we obtain
\begin{equation}
\rho= \rho_{1} a^{-2 \gamma}, \label{density}
\end{equation}
where the constant $\rho_{1}$ is the energy density of the
universe at $t=1$.

Since the particle horizon is defined by the distance covered by
the light cone emitted at the Planck time $t=1$:
\begin{equation}
\label{hor} \chi(t)=\chi_{1}+ \int^{t}_{1} \frac{dt'}{a(t')}.
\end{equation}
Notice that $\chi(t=1)=\chi_{1}\sim 1$, since the physical
distance $d_{h}$ is given by $a(t)\chi(t)$ and it must of the
order of $l_{P}$ at the Planck time $t=1$. The assumed initial
conditions implies that the range of the coordinate $\chi$ is $0
< \chi < \pi$, for $\kappa=1$, and $0 < \chi < \infty$, for
$\kappa=0,-1$. In the formulation of HP by Fischler and Susskind,
the entropy contained within a volume coordinate of size $R_{H}$
should not exceed the area of the horizon in Planck units. In
terms of the (constant) comoving entropy density $\sigma$
\begin{equation}
\label{FS}
 \sigma R_{H}^{d}< (aR_{H})^{d-1}.
 \end{equation}

Due to its simplicity, it is easy to treat the flat ($k=0$) models
in $D=d+1$ dimensions, where $d$ is the spatial dimension.
Introducing Eq.(\ref{density}) in Eq.(\ref{fe1}) and solving for
a(t), we find (for $\gamma \neq 0$)
\begin{equation}
\label{agama}
 a(t) \sim t^{2/d\gamma}.
\end{equation}
The particle horizon is given by
\begin{eqnarray}
\chi (t)  & \sim & t^{1-2/d\gamma} , \mbox{ if
$\mbox{$\gamma$}\neq 2/d$},
 \label{chi1} \\ \chi (t) & \sim & 1+\ln t , \mbox{ if
$\mbox{$\gamma$}=2/d$}. \label{chi2}
\end{eqnarray}
The solutions given by Eq. (\ref{chi1}) and Eq.(\ref{chi2})for the
comoving size show that at the Planck time its value it is not
zero. The holographic bound is given,in terms of $\chi(t)$, by
\begin{equation}
\label{hp} \frac{S(t)}{A}= \sigma \frac{\chi (t)}{a(t)^{d-1}}< 1.
\end{equation}
Introducing the expression for $\chi (t)$ in Eq.(\ref{hp}) we
obtain
\begin{eqnarray}
\frac{S(t)}{A}& = & \sigma t^{1-2/\gamma} , \mbox{ if
$\mbox{$\gamma$}\neq 2/d$},
 \label{SA1} \\ \frac{S(t)}{A}& = & \sigma \frac{1+\ln t}{t^{d-1}}, \mbox{ if
$\mbox{$\gamma$}=2/d$}. \label{SA2}
\end{eqnarray}

The assumption that the comoving size that at the Planck time is
not zero is crucial in the case $\gamma=2/d$, since otherwise
$S(t)/A = 0$ at the beginning.

The last two equations indicates that in $d$ dimensions the ratio
$S/A$ does not increase in time for $0 <\gamma < 2$.  If the
constraint $S/A $ was satisfied at the Planck time, later on it
will be satisfied even better.It is easy to prove that the cases
$\gamma=0$ and $\gamma=2$, which correspond to a de Sitter
universe and a universe filled with stiff matter, respectively,
also satisfies the holographic bound. Thus, for any dimension,
the FS's condition is satisfied by flat universes filled with
matter satisfying DEC. Of course this include the cases of
universes filled with dust or radiation. The universe is
inflationary if $\gamma < 2/d$. In $2+1$ dimensions, this implies
that universes filled with any exotic fluids with negative
pressure are inflationary. The above results constitute a
generalization of the previous one found in \cite{Wang}, which
are related with non exotic fluid only. In $3+1$ dimensions,
inflationary universes require, in order, to have mechanism of
matter creation that leads to the standard model, strongly
nonadiabatic process as the reheating \cite{Kaloper}. Since, 2+1
dimensional gravity is a theoretical model not related with our
observable universe, we do not consider relevant to further
discuss these process. Only the classical solution is considered,
which represents an universe inflating for ever.

In the following we shall discuss the not flat cases
corresponding to a closed and open $2+1$ universes. In these cases
the area of the event horizon is given by
\begin{equation}
\label{area}
 A(\chi) = 2 \pi a f_{\kappa}(\chi),
 \end{equation}
and the volume bounded by $A(\chi)$ is given by
\begin{equation}
\label{volumen}
 V(\chi) = 2 \pi \int d\chi f_{\kappa}(\chi).
 \end{equation}
The condition (\ref{FS}) can be expressed in the following form
\begin{equation}
\label{hp} \frac{S(\chi)}{A}= \sigma \frac{V(\chi)}{A(\chi)}<1.
\end{equation}

Evaluating the volume and the area for a close and open universe,
we obtain

\begin{eqnarray}
\frac{S(\chi)}{A} & = & \sigma \frac{1-\cos\chi}{a(\chi)
\sin\chi} \qquad \mbox{ for $\mbox{$\kappa$}=1$}\label{SK1},
\\
\frac{S(\chi)}{A} & = & \sigma \frac{\cosh\chi-1}{a(\chi)
\sinh\chi} \qquad \mbox{ for $\mbox{$\kappa$}=-1$}\label{SK-1}.
\end{eqnarray}

We emphasize again that $\chi$ is not zero at the initial time.

The condition $S/A < 1$ implies that HP is satisfied if
\begin{equation}
a(\chi) > \left\{
\begin{array}{cl}
 \displaystyle \sigma (1-\cos\chi)(\sin\chi)^{-1}
& \qquad  \mbox{if $\kappa=1$}, \\
 \mbox{$\sigma (\cosh\chi-1)(\sinh\chi)^{-1}$} & \qquad \mbox{if
 $\kappa=-1$}.
\end{array} \right. \label{SK1}
\end{equation}

The solutions for the scale factor in terms of $\chi$, for the
closed and open universes, are given by
\begin{equation}
a^{\gamma -1}(\chi) \sim \left\{
\begin{array}{cl}
 \displaystyle \sin ((\gamma -1)\chi)
 & \qquad  \mbox{if $\kappa=1$}, \\
 \mbox{$\sinh ((\gamma -1)\chi)$} & \qquad \mbox{if $\kappa=-1$}.
\end{array} \right. \label{aclose}
\end{equation}
These solutions are valid for $1< \gamma \leq 2$, i. e., $p>0$.
In the case $\kappa=-1$, the solution given in Eq.(\ref{aclose})
is valid if $\gamma \neq 1$.  If $\chi \rightarrow \pi$ then $a
\rightarrow 0$ and the bound in Eq.(\ref{SK1}) for a closed
universe is not satisfied. For an open universe the bound in
Eq.(\ref{SK-1}) is satisfied.

For a dust filled universe,\,\,$\gamma = 1$,\,\,$p=0$,
Eq.(\ref{fe1}) and Eq.(\ref{density}) leads to the solution
\begin{equation}
\label{flat} a(t) = 1+ {\sqrt{\alpha_{1}-k}}\,t,
\end{equation}
where $\alpha_{\gamma}=2\pi G \rho_{1}^{(\gamma)}$ and
$\rho_{1}^{(\gamma)}$ is the initial density for a $\gamma$-fluid.
This solution implies that the universe will always expand,
regardless of the value of $k$, if $\alpha_{1} -k >0$. In terms
of $a(\chi)$ the solution is
\begin{equation}
a(\chi) = \exp (\sqrt{(\alpha_{1} -k)}\,\chi), \label{chi}
\end{equation}
with the restriction $\alpha_{1} > 1$ for closed universes. The
inconsistency with HP remains for closed universes. Notice that
the result found in \cite{Wang} is obtained using the expression
for the horizon corresponding to a flat universe, which
constitute a lack of consistence. The conclusion is that for
closed universes filled with fluids with $1 < \gamma \leq 2$ the
holographic bound is not satisfied.

%%%%%%%%%%%%%%%%%%%%%%%%%%%%%%%%%%%%%%%%%%%%%%%%%%%%%%%%%%%%%%%%%%%%%%%%%%
%   FLUIDOS EXOTICOS con comentario sobre las condiciones de energia
%%%%%%%%%%%%%%%%%%%%%%%%%%%%%%%%%%%%%%%%%%%%%%%%%%%%%%%%%%%%%%%%%%%%%%%%%%

Previous results in $3+1$ dimensions indicates that closed
universes may satisfy the holographic bound if two fluids are
included, one of them with negative pressure. From the point of
wiew of the energy conditions, negative pressure cannot satisfy
the {\it strong energy condition} (SEC). In the $3+1$ dimensional
case SEC implies $\rho + 3p\,>\,0$. Nevertheless, in $2+1$ SEC
implies that the pressure must be positive, $p\,>\,0$
\cite{Barrow}. Therefore, in $2+1$ dimensions if $\rho>0$ and
$p\,>\,0$ then \cite{Visser}
\begin{equation}
\ddot{a}\,<\,0. \label{acelera}
\end{equation}
The corresponding cosmological solutions must be non
inflationary, independent of whether the Universe is open, flat,
or closed. SEC can be violated by a positive cosmological
constant or by a fluid with negative pressure \cite{Visser1}.

In the $2+1$ dimensional case, since the acceleration depends only
on the pressure, an inflationary solution for the scale factor is
obtained if the total pressure is negative, which is the case if
the matter content of the universe is only one fluid with
negative pressure. For a universe filled with dust and other
exotic fluid, the scale factor is always accelerating. Solutions
with a late-time accelerated expansion are found for universes
filled with two fluids, one with $1\,<\,\gamma \leq 2$, and the
other negative pressure. We will analyze below cosmological
solutions with exotic fluids. It is not difficult to find exact
solutions to the Einstein's equation
\begin{equation}
\dot{a}^{2}  = -\kappa +  2 \pi G
\rho_{1}^{(\gamma)}a^{2(1-\gamma)},
 \label{exotic1}
\end{equation}
for the cases with $\gamma = 1/2$ and $\gamma = 0$. We first show
the solution for $\gamma = 1/2$. For a closed universe, $a(\chi)$
is given by
\begin{equation}
a(\chi)  =  \frac{1}{\alpha_{1/2}}\frac{1}{\cos^{2}(\chi/2 +
\arctan (\sqrt{\alpha_{1/2}-1}))}.
 \label{exotic1}
\end{equation}
The solution corresponding to an open universe is
\begin{equation}
a(\chi)  = \frac{4\delta}{\alpha_{1/2}}\frac{\exp \chi}
{(\delta-\exp \chi)^{2}} ,
 \label{exotic2}
\end{equation}
where $\delta \equiv
\frac{\sqrt{1+\alpha_{1/2}}+1}{\sqrt{1+\alpha_{1/2}}-1}$. From
the condition given in Eq.(\ref{SK1}) it is easy to see that HP is
satisfied. In terms of $a(t)$ the solution is (see \cite{Cruz})
\begin{equation}
a(t)  =
1+\sqrt{\alpha_{1/2}-k}\,\,(t-1)+\frac{1}{4}\alpha_{1/2}(t-1)^{2}.
 \label{exotict}
\end{equation}

Let us consider the case $\gamma =0$. For a closed universe,
$a(\chi)$ is given by
\begin{equation}
a(\chi)  =  \frac{\eta}{\sin(\arcsin\eta - \chi)},
\label{exotic11}
\end{equation}
where $\eta \equiv (\alpha_{0})^{-1/2}$ and we fix $\eta$
throughout $\arcsin \eta = \pi/4$, in order to have a physical
solution. This is equivalent to choose the value of the initial
density. The solution $a(t)$ is given by
\begin{equation}
a(t)  =  \frac{\eta}{2}\left(\beta \exp ((t-1)/\eta)+(\beta \exp
(( t-1)/\eta))^{-1}\right), \label{exotict11}
\end{equation}
where $\beta=\eta^{-1}(1+(1-\eta^{2d})^{1/2})$.

The solution corresponding to an open universe is
\begin{equation}
a(\chi)  =  \frac{\eta}{\sinh(\sinh^{-1}\eta - \chi)},
 \label{exotic22}
\end{equation}
and in terms of $t$ we obtain
\begin{equation}
a(t)  = \eta \sinh(\sinh^{-1}\eta^{-1} + (t-1)/\eta),
 \label{exotict22}
\end{equation}

From the condition given in Eq.(\ref{SK1}) it is easy to see that
HP is satisfied. As a conclusion, we can say that exist
cosmological solutions of closed universes filled with exotic
fluids that satisfy HP.

%%%%%%%%%%%%%%%%%%%%%%%%%%%%%%%%%%%%%%%%%%%%%%%%%%%%%%%%%%%%%%%%%%%
%%%%%%%%%%%%%%%%%%%%%%%%%%%%%%%%%%%%%%%%%%%%%%%%%%%%%%%%%%%%%%%%%%%
%%%%%%%%%%%%%%%%%%%%%%%%%%%%%%%%%%%%%%%%%%%%%%%%%%%%%%%%%%%%%%%%%%%
%    SOLUCION CON DOS FLUIDOS
%
%%%%%%%%%%%%%%%%%%%%%%%%%%%%%%%%%%%%%%%%%%%%%%%%%%%%%%%%%%%%%%%%%%%
%%%%%%%%%%%%%%%%%%%%%%%%%%%%%%%%%%%%%%%%%%%%%%%%%%%%%%%%%%%%%%%%%%%
%%%%%%%%%%%%%%%%%%%%%%%%%%%%%%%%%%%%%%%%%%%%%%%%%%%%%%%%%%%%%%%%%%%

It is interesting to discuss universes filled with two fluid,
since in $3+1$ dimensions the solutions found in \cite{Rama} for
a closed universe satisfied HP. In this case, exist two fluids,
where one of them has an equation of state, $p= w\rho$, with $w <
-1/3$. We had conclude above that solution with only one exotic
fluid with negative pressure satisfy HP. If we consider universes
with two fluids one of them possessing zero pressure, we expected
that, since the total pressure continue being negative, an
inflationary solution that may also satisfy HP. We discuss bellow
two cases with two fluids where it is possible to find exact
solutions. The first case corresponds to dust plus and exotic
fluid with $\gamma =1/2$. Solving the field equations, we obtain
for $a(\chi)$
\begin{equation}
a(\chi)  = \frac{\xi(k)}{\alpha_{1/2}}\left (\left (\frac{\lambda
+ \exp \xi(k)\chi}{\lambda - \exp \xi(k)\chi}\right )^{2}-1\right
),
 \label{twofluid}
\end{equation}
where $\xi(k)= \alpha_{1}-k$ and
$\lambda=\frac{\sqrt{\xi(k)+\alpha_{1/2}}+\sqrt{\xi(k)}}
{\sqrt{\xi(k)+\alpha_{1/2}}-\sqrt{\xi(k)}}$. It is
straightforward to show that flat and open universes verify HP.
For the closed case, if $\chi \rightarrow \ln \lambda
/\xi(1)<\pi$ is straightforward to verify HP. This case can be
reduced, by a suitable rescaling, to a one fluid case. Although
mathematically equivalent they are physically distinguishable.
The solution for $a(t)$ is given in Eq.(\ref{exotict}), except
that the coefficient of $t-1$ change to
$\sqrt{\alpha_{1/2}-k+\alpha_{1}}$.

The second case corresponds to a fluid with $\gamma =3/2$
(positive pressure) and exotic fluid with $\gamma =1/2$. We
obtain for a closed universe that $a(\chi)$ is given by the
following expression
\begin{equation}
a(\chi)  = \frac{1}{2\alpha_{1/2}}\left (\frac{C + \exp
\chi/\sqrt{2}}{C - \exp\chi/\sqrt{2}}\right )^{2},
 \label{twofluid1}
\end{equation}
where $C=\frac{\sqrt{2+\alpha_{1/2}}+1}{\sqrt{2\alpha_{1/2}}-1}$.
For this case we have choose $4\alpha_{3/2}\alpha_{1/2}=1$. If
$\chi \rightarrow \sqrt{2}\ln C<\pi$, HP is verified. In this
case is possible to obtain only an implicit solution for $a$.

%%%%%%%%%%%%%%%%%%%%%%%%%%%%%%%%%%%%%%%%%%%%%%%%%%%%%%%%%%%%%%%%%%
%%%%%%%%%%%%%%%%%%%%%%%%%%%%%%%%%%%%%%%%%%%%%%%%%%%%%%%%%%%%%%%%%%%
%%%%%%%%%%%%%%%%%%%%%%%%%%%%%%%%%%%%%%%%%%%%%%%%%%%%%%%%%%%%%%%%%%%
%    SOLUCION CON lambda negativo
%
%%%%%%%%%%%%%%%%%%%%%%%%%%%%%%%%%%%%%%%%%%%%%%%%%%%%%%%%%%%%%%%%%%%
%%%%%%%%%%%%%%%%%%%%%%%%%%%%%%%%%%%%%%%%%%%%%%%%%%%%%%%%%%%%%%%%%%%
%%%%%%%%%%%%%%%%%%%%%%%%%%%%%%%%%%%%%%%%%%%%%%%%%%%%%%%%%%%%%%%%%%%

Let us discuss universes which contain a negative cosmological
constant. A universe filled with a fluid satisfying DEC and
$-\Lambda < 0$, collapse, independently of the curvature of the
hipersurface of homogeneity. HP is violated for these universes
following the formulations of FS and Bak and Rey ~\cite{Bak}.
Also the Bousso's proposal is not valid if a negative
cosmological constant is present. In $2+1$ dimensions the
situation is quite similar and HP is violated before the time of
collapse. We consider the simple case of $k=0$, and $1<\gamma
\leq 2$. The field equation is given by
\begin{equation}
\label{lambdaneg}
 (\frac{\dot{a}}{a})^{2} =
\frac{\alpha_{\gamma}}{a^{2\gamma}}-\Lambda.
\end{equation}
For this case $\chi(a)$ is given by
\begin{equation}
\label{hor} \sqrt{\alpha_{\gamma}}\chi(a)= \int^{a}_{1}
x^{\gamma-2}\left (1-\frac{\Lambda x^{2\gamma}}
{\alpha_{\gamma}}\right)^{-1/2}dx.
\end{equation}
Defining the new variable $\zeta (x) =
(\Lambda/\alpha_{\gamma})^{1/2}x^{\gamma}$, we obtain
\begin{equation}
\label{zeta} \sqrt{\alpha_{\gamma}}\chi(a)=
\int^{\zeta(a)}_{\zeta(1)}\epsilon^{-1/\gamma}\left
(1-\epsilon^{2}\right)^{-1/2}d\epsilon.
\end{equation}
At the turning point, $a_{t}$, $\dot{a}=0$, i. e., $\zeta
(a_{t})=1$ and consistently
$a_{t}=(\frac{\alpha_{\gamma}}{\Lambda})^{1/2\gamma}$. Then $\chi
(a_{t})$ is given by
\begin{equation}
\label{zetat} \sqrt{\alpha_{\gamma}}\chi(a_{t})=
\int^{1}_{\zeta(1)}\epsilon^{-1/\gamma}\left
(1-\epsilon^{2}\right)^{-1/2}d\epsilon.
\end{equation}
In order to evaluate the above integral we rewrite it as
\begin{equation}
\label{zet-a} \int^{1}_{\zeta(1)}\epsilon^{-1/\gamma}\left
(1-\epsilon^{2}\right)^{-1/2}d\epsilon =
\int^{1}_{0}\epsilon^{-1/\gamma}\left
(1-\epsilon^{2}\right)^{-1/2}d\epsilon -
\int^{\zeta(1)}_{0}\epsilon^{-1/\gamma}\left
(1-\epsilon^{2}\right)^{-1/2}d\epsilon.
\end{equation}
Evaluating these integrals we obtain
\begin{eqnarray}
\label{zetat} \sqrt{\alpha_{\gamma}}\chi(a_{t}) & = &
\frac{1}{2\gamma}\left
(\frac{\alpha_{\gamma}}{\Lambda}\right)^{\frac{\gamma-1}{2\gamma}}B\left
(\frac{\gamma-1}{2\gamma},\frac{1}{2}\right)
 \\  & - & \left
(\frac{1}{\gamma-1}+\frac{1}{2(3\gamma-1)}\frac{\Lambda}{\alpha_{\gamma}}
+\frac{3}{8(5\gamma-1)}\left (
\frac{\Lambda}{\alpha_{\gamma}}\right)^{2\gamma}+ ...\right),
\end{eqnarray}
where $B(x,y)$ is the Euler's beta function, which correspond to
the first integral in the r.h.s. of Eq. (\ref{zet-a}) and the
expansion in powers of $\frac{\Lambda}{\alpha_{\gamma}}$
correspond to the evaluation of the second integral in the r.h.s.
of equation (\ref{zet-a}). If we assume that
$a_{t}=(\frac{\alpha_{\gamma}}{\Lambda})^{1/2\gamma}a_{0}\gg 1$ (
we have restored $a_{0}$ in order to clarify the discussion),
which correspond a universe that can evolve to a size many orders
of magnitude greater than the Planck scale, the expansion in
powers of $\frac{\Lambda}{\alpha_{\gamma}}$ is negligible.

Then, from Eq.(\ref{hp})the quotient $S/A$ at the turning point
results
\begin{equation}
\label{turning} \frac{S}{A}(a_{t})=
\frac{\sigma}{2\gamma\alpha_{\gamma}^{1/2}}B\left (
\frac{\gamma-1}{2\gamma},\frac{1}{2}\right)\left
(\frac{\Lambda}{\alpha_{\gamma}}\right)^{\frac{2-\gamma}{2\gamma}}.
\end{equation}
It is direct to show that $\frac{S}{A}(a_{t})$ satisfy HP at the
turning point since $\frac{\Lambda}{\alpha_{\gamma}}< 1$, as it
was mentioned above. The particle horizon defined by $L=a\chi$
satisfy near the collapse point the relation $L_{c}\sim
2L_{t}/a_{t}$, where $L_{c}$ and $L_{t}$ are the particle horizon
near the collapse point, $a_{c}$, and at the turning point
respectively; $a_{c}\sim a_{0}$. Then $\frac{S}{A}$ near the final
stage of collapse is given by
\begin{equation}
\label{collapse} \frac{S}{A}\sim
\frac{\sigma}{\gamma\alpha_{\gamma}^{1/2}}B\left (
\frac{\gamma-1}{2\gamma},\frac{1}{2}\right)\left
(\frac{\Lambda}{\alpha_{\gamma}}\right)^{\frac{-(\gamma-1)}{2\gamma}}.
\end{equation}
In this case $\frac{S}{A}\gg 1$ at collapse point. Therefore,
$\frac{S}{A}$ reaches the unity at some time after the turning
point, and HP is not satisfied thereafter.

In 2+1 dimensions is possible to verify that this occur also for
$k\neq0$, using explicitly the cosmological solutions found in
\cite{Cruz} for $a(t)$. The solution $a(\chi)$ for dust,
$\gamma=1$, and for all $k$, is given by
\begin{equation}
a(\chi)  = \frac{2\theta \left
(\theta+\sqrt{\theta^{2}-1}\right)}{\exp\theta\sqrt{\Lambda}\chi
+\left(\theta+\sqrt{\theta^{2}-1}\right)^{2}\exp -
\theta\sqrt{\Lambda}\chi },
 \label{polvok}
\end{equation}
where $\theta^{2}=(\alpha_{1}-k)/\Lambda>0$. It is straightforward
to verify that HP is violated in the future.

As a conclusion, homogeneous and isotropic universes, even with a
small contribution of a negative density of energy, collapse and
not satisfy HP, independently of the type of curvature.

In order to clarify the behavior of the found solutions we will
analyse whether the curvature invariants remain finite until the
evolution enters the Planckian regime. For FRW models filled with
fluids obeying a $\gamma$-law equation of state and a cosmological
constant, the invariant $R_{\mu\nu}R^{\mu\nu}$ is given by
\begin{equation}
R_{\mu\nu}R^{\mu\nu}=12 \Lambda^{2}+\frac{8\alpha_{\gamma}\Lambda
(2\gamma-3)}{a^{2\gamma}}+\frac{2\alpha_{\gamma}^{2}(3\gamma^{2}-8\gamma
+6 )}{a^{4\gamma}}.
 \label{invarian}
\end{equation}
It is easy to see that this invariant remains finite when
$a\rightarrow 1$ for all solutions found. The singularity is at
$a=0$, as in the $3+1$ dimensional models. The other invariants
of interest exhibits the same behavior.

In summary, we have discussed the holographic principle in $2+1$
universes filled with one or two fluids that obey a $\gamma$-law
equation of state. Our principal objective was to clarify, within
the FS's proposal for the holography in cosmology, the behavior of
closed models, that in previous investigations appear not satisfy
HP no matter whether the universe is composed. Our results show
that HP is maintained if the universe is filled with exotic
matter, with negative pressure. We also generalize the results
for flat universes, showing that HP is satisfied if the fluid
obey DEC.

Violations of FS's proposal by physically reasonable universes,
such as the closed one, have lead to other approaches that, for
example, consider bound the entropy inside space-like regions. In
the approach outlined by Veneciano \cite{Veneziano}, the
holographic bound on the entropy of the observable part of the
universe is related to the theory of black holes, suggesting that
the entropy of the interior of a domain of size $H^{-1}$ cannot
be greater than the entropy of a black hole of a similar radius.
These ideas are based on the Bekenstein entropy bound for any
isolated physical system \cite{Bound}. The constrains derived
from this approach do not ruled out universes with a negative
cosmological constant ( a desirable result, since many
applications of the HP have been studied in AdS space) and do not
impose any additional constrains on inflationary cosmology.
Nevertheless, in $2+1$ universes this approach must be applied
taking account that black hole solutions exist only when a
negative cosmological constant is considered.

%%%%%%%%%%%%%%%%%%%%%%%%%%%%%%%%%%%%%%%%%%%%%%%%%%%%%%%%%%%%
%  Comentarios
%
%%%%%%%%%%%%%%%%%%%%%%%%%%%%%%%%%%%%%%%%%%%%%%%%%%%%%%%%%%%%%%%%%%%

\section*{Acknowledgements}
 The authors acknowledge the referee for useful suggestions in
 order to improve the presentation of the results of this paper.
 This work was supported by USACH-DICYT under Grant N$^0$ 04-0031CM(NC)
 and by CONICYT through Grant N$^0$ 2990037(SL).

\end{document}